\documentstyle[amssymb,aps]{revtex}

\tightenlines

\begin{document}
\title{Prospects for photon blockade in four level systems in the N configuration
with more than one atom.}
\author{Andrew D. Greentree$^{\ast }$, John A. Vaccaro$\dagger $, Sebasti\'{a}n R.
de Echaniz$^{\ast }$, Alan V. Durrant$^{\ast }$, Jon P. Marangos$\ddagger $}
\address{$^{\ast }$Quantum Processes Group\\
Department of Physics,\\
The Open University,\\
Milton Keynes MK7 6AA,\\
United Kingdom\\
$^{\dagger }$Department of Physics and Astronomy,\\
The University of Hertfordshire,\\
Hatfield AL10 9AB,\\
United Kingdom\\
$^{\ddagger }$Laser Optics and Spectroscopy Group,\\
Blackett Laboratory,\\
Imperial College,\\
London SW7 2BZ,\\
United Kingdom}
\date{25 February 2000}
\maketitle

\begin{abstract}
We show that for appropriate choices of parameters it is possible to achieve
photon blockade in idealised one, two and three atom systems. \ We also
include realistic parameter ranges for rubidium as the atomic species. \ Our
results circumvent the doubts cast by recent discussion in the literature
(Grangier {\it et al} Phys. Rev Lett. {\bf 81}, 2833 (1998), Imamo\u{g}lu 
{\it et al} Phys. Rev. Lett. {\bf 81}, 2836 (1998)) on the possibility of
photon blockade in multi-atom systems.
\end{abstract}

\section{Introduction}

Recent work by Imamo\u{g}lu {\it et al}\cite{bib:Imamoglu1997}{\it \ }%
suggested a promising scheme for observing photon blockade in a highly
non-linear cavity, where the change in the Kerr nonlinearity due to single
photon effects was enough to make the cavity non-resonant with modes of more
than one photon. As described in their work, such a device could work as a
single-photon turnstile, similar to that realised recently by Kim {\it et al}
\cite{bib:Kim1999} in a semiconductor junction, which might be useful for
quantum computation and the generation of non-classical light fields. The
work by Imamo\u{g}lu {\it et al} was based on the use of the adiabatic
elimination procedure and later studies\cite
{bib:Imamoglu1998,bib:Grangier1998} showed that the breakdown of the
procedure in the high dispersion limit leads to prohibitive restrictions on
the parameter space where photon blockade could be observed in a multi-atom
system. \ Further to this, Werner and Imamo\u{g}lu\cite{bib:Werner1999} and
Rebi\'{c} {\it et al}\cite{bib:Rebic1999} suggested that these problems
could be overcome by using a system with a single atom. \ The question of
observing photon blockade in multi-atom systems is of more than just
theoretical interest. \ Possible schemes for observing photon blockade
depend on directing a low flux atomic beam through a high finesse cavity and
this implies that there will be an uncertainty in the number of atoms
present in the cavity at any given time. \ We show that the photon blockade,
in certain circumstances, remains strong even for a fluctuation in the atom
number. \ Recent work\cite{bib:Werner1999,bib:RebicPrePrint} has highlighted
the importance of employing a mutual detuning between the cavity and the
semiclassical coupling field to shift the many atom degenerate state out of
resonance. \ Our work builds on this idea and presents a detailed
atom-cavity dressed state calculation showing the parameter regimes which
offer the best prospects for the observations of photon blockade in one, two
and three atom systems. \ We show that current technology should allow the
construction of a system exhibiting photon blockade. \ We suggest that
photon blockade can be best realised, not in a MOT as originally envisaged
by Imamo\u{g}lu {\it et al}\cite{bib:Imamoglu1997}, but by sending a low
atom current beam through a cavity of the style realised by Hood {\it et al} 
\cite{bib:Hood1998} and M\"{u}nstermann {\it et al}\cite{bib:Munstermann}.

Photon blockade in a cavity can be explained quite simply. \ An external
field drives a cavity that is resonant when there are zero or one photon in
the cavity and non-resonant for two (or more) photons. \ This can be
achieved by introducing a medium into the cavity, exhibiting a large Kerr
non-linearity which alters the refractive index as a function of the
intensity. \ In general, one would expect to find large non-linearities in
systems exhibiting electromagnetically induced transparency \cite{bib:EIT}.

\section{Dressed states of the atom-cavity system}

Consider a four level atom as depicted in figure 1. \ The energy levels are
labeled in order of increasing energy as $\left| a\right\rangle $, $\left|
b\right\rangle $, $\left| c\right\rangle $ and $\left| d\right\rangle $ with
associated energies of $\hbar \omega _{a}$, $\hbar \omega _{b}$, $\hbar
\omega _{c}$ and $\hbar \omega _{d}$ respectively and transition frequencies 
$\omega _{\alpha \beta }=\omega _{\alpha }-\omega _{\beta }$ where $\alpha
,\beta =a,b,c,d$. \ The $\left| b\right\rangle -\left| c\right\rangle $
transition is driven by a strong classical coupling field, with frequency $%
\omega _{\text{class}}$, Rabi frequency $\Omega $ , detuned from the $\left|
b\right\rangle -\left| c\right\rangle $ transition by an amount $\delta
_{cb}=\omega _{\text{class}}-\omega _{cb}$. \ The atoms are in a cavity with
resonance frequency $\omega _{\text{cav}}$ which is detuned from the $\left|
a\right\rangle -\left| c\right\rangle $ transition by an amount $\delta
_{ca}=\omega _{\text{cav}}-\omega _{ca}$, detuned from the $\left|
b\right\rangle -\left| d\right\rangle $ transition by an amount $\Delta
=\omega _{\text{cav}}-\omega _{db}$ and not interacting with the $\left|
b\right\rangle -\left| c\right\rangle $ transition. \ The detunings $\delta
_{cb}$ and $\delta _{ca}$ are set equal to ensure that the $\left|
a\right\rangle -\left| b\right\rangle $ transition is driven by a two photon
resonance. We therefore define the mutual detuning, $\delta =\delta
_{cb}=\delta _{ca}$. \ The cavity is driven by an additional classical field
with frequency $\omega _{e}=\omega _{\text{cav}}$ and power, $P$. \ We
analyse the effect of this field by examining the dressed states of the
atom-cavity system. \ This configuration of fields and atomic levels is
called the N configuration and has been considered previously \cite
{bib:Imamoglu1997,bib:Werner1999,bib:Rebic1999,bib:Schmidt1996}. \ The
cavity linewidth is $\Gamma _{\text{cav}}$. \ The atom-cavity mode coupling
is $g_{\alpha _{1}\alpha _{2}}=\left( \omega _{\alpha _{1}\alpha
_{2}}/2\hbar \epsilon _{0}V_{\text{cav}}\right) ^{1/2}\mu _{\alpha
_{1}\alpha _{2}}$ where $\alpha _{1}$ and $\alpha _{2}$ correspond to atomic
levels, $\mu _{\alpha _{1}\alpha _{2}}$ is the electric dipole moment of the
transition, $V_{\text{cav}}$ is the cavity volume and $\epsilon _{0}$ is the
permittivity of free space. \ The Hamiltonian for the system with $N$ atoms
in the frame rotating at the cavity resonance frequency in the rotating-wave
approximation is\cite{bib:Werner1999} 
\begin{eqnarray}
\frac{\widehat{{\cal H}}}{\hbar } &=&-i\tilde{\Gamma}_{c}\sum_{j=1}^{N}%
\widehat{\sigma }_{cc}^{j}-i\widetilde{\Gamma }_{d}\sum_{j=1}^{N}\widehat{%
\sigma }_{dd}^{j}+\sum_{j=1}^{N}\Omega \left( \widehat{\sigma }_{cb}^{j}+%
\widehat{\sigma }_{bc}^{j}\right)  \nonumber \\
&&+\sum_{j=1}^{N}g_{ac}\left( \widehat{a}\widehat{\sigma }_{ca}^{j}+\widehat{%
a}^{\dagger }\widehat{\sigma }_{ac}^{j}\right) +\sum_{j=1}^{N}g_{bd}\left( 
\widehat{a}\widehat{\sigma }_{db}^{j}+\widehat{a}^{\dagger }\widehat{\sigma }%
_{bd}^{j}\right) -i\Gamma _{\text{cav}}\widehat{a}^{\dagger }\widehat{a}.
\label{eq:Hamiltonian}
\end{eqnarray}
where $\tilde{\Gamma}_{c}=\Gamma _{c}+i\delta $, $\widetilde{\Gamma }%
_{d}=\Gamma _{d}+i\Delta $, $\Gamma _{\alpha }$ is the decay rate from
atomic state $\left| \alpha \right\rangle $, $\widehat{a}$ $\left( \widehat{a%
}^{\dagger }\right) $ is the cavity photon annihilation (creation) operator
and $\widehat{\sigma }_{\alpha _{1}\alpha _{2}}^{j}$ is the atomic operator $%
\left| \alpha _{1}\right\rangle \left\langle \alpha _{2}\right| $ acting on
atom $j$.

We find the dressed states by diagonalizing $\widehat{{\cal H}}$. \
Fortunately, $\widehat{{\cal H}}$ is block diagonal in the bare state basis.
\ The $n$th block can be identified by starting with the state $\left|
a,a,\ldots ,n\right\rangle $, representing all the atoms in state $\left|
a\right\rangle $ and the cavity field in the $n$ photon state $\left|
n\right\rangle $, and finding the closed set of states coupled to $\left|
a,a,\ldots ,n\right\rangle $ by $\widehat{{\cal H}}$. \ Diagonalizing this
block gives the $n$ quanta manifold of dressed states.

We first consider the case of a single atom in the cavity. \ The zero quanta
manifold consists solely of the state $\left| a,0\right\rangle $. \ The one
quantum manifold is spanned by the states $\left| a,1\right\rangle $, $%
\left| b,0\right\rangle $ and $\left| c,0\right\rangle .$ The corresponding
block of $\widehat{{\cal H}}/\hbar $ \ can be written in matrix form in this
basis as 
\begin{equation}
\frac{{\cal H}_{1}^{\left( 1\right) }}{\hbar }=\left[ 
\begin{array}{ccc}
-i\Gamma _{\text{cav}} & 0 & g_{ac} \\ 
0 & 0 & \Omega \\ 
g_{ac} & \Omega & -i\tilde{\Gamma}_{c}
\end{array}
\right]  \label{eq:MatrixForm:1Atom1Photon}
\end{equation}
where the superscript on ${\cal H}$ refers to the number of atoms and the
subscript to the number of quanta in the system. \ In order to simplify the
expressions for eigenvalues and eigenvectors, we assume $\Gamma _{\text{cav}%
}=0$. \ The figures which follow, however, have been generated using
non-zero values of $\Gamma _{\text{cav}}$. \ Diagonalising the matrix ${\cal %
H}_{1}^{\left( 1\right) }/\hbar $ with $\Gamma _{\text{cav}}=0$ gives the
dressed state energies 
\begin{eqnarray*}
{\cal E}_{+} &=&\left( -i\tilde{\Gamma}_{c}+\sqrt{-\tilde{\Gamma}%
_{c}^{2}+4\left( \Omega ^{2}+g_{ac}^{2}\right) }\right) /2 \\
{\cal E}_{0} &=&0 \\
{\cal E}_{-} &=&\left( -i\tilde{\Gamma}_{c}-\sqrt{-\tilde{\Gamma}%
_{c}^{2}+4\left( \Omega ^{2}+g_{ac}^{2}\right) }\right) /2
\end{eqnarray*}
and corresponding dressed states $\left| D_{+}\right\rangle $, $\left|
D_{0}\right\rangle $ and $\left| D_{-}\right\rangle $ respectively. \ In
this form, the real part of the eigenstate corresponds to the state energy
and the imaginary part to the width of the state. \ These eigenstates form
the well known Mollow triplet \cite{bib:Mollow1969} and are presented in
figure 2(a) as a function of the scaled mutual detuning, $\delta /\Omega $,
with $\Gamma _{c}/\Omega =0.1$, $\Gamma _{\text{cav}}/\Omega =0.01$ and $%
g_{ac}/\Omega =1$. \ It is important to express the form of the central
dressed state, $\left| D_{0}\right\rangle $ which is 
\[
\left| D_{0}\right\rangle =\frac{\Omega }{\sqrt{\Omega ^{2}+g_{ac}^{2}}}%
\left| a,1\right\rangle -\frac{g_{ac}}{\sqrt{\Omega ^{2}+g_{ac}^{2}}}\left|
b,0\right\rangle . 
\]

Photon blockade will occur in this dressed-state picture when the cavity
driving field resonantly couples the zero to one quantum manifolds, and only
weakly couples the one and two quanta manifolds. \ We can gauge the extent
of these couplings by treating each transition driven by the cavity driving
field as a separate, closed two-state system. \ This approach will break
down when multiple states are excited simultaneously. \ However, in
situtations where the photon blockade effect occurs, the number of dressed
states that are significantly occupied will be minimal and our two-state
model should give a reasonably accurate picture of the degree of excitation
of each transition.

The effect of the cavity driving field on the cavity-atom system can be
treated by including the additional term on the right hand side of equation 
\ref{eq:Hamiltonian} 
\[
\hbar \beta (\hat{a}+\hat{a}^{\dagger }) 
\]
where $\beta =\sqrt{P\Gamma _{\text{cav}}T^{2}/\left( 4\hbar \omega _{\text{%
cav}}\right) }$ is the external field-cavity mode coupling strength for a
cavity mirror transmittance of $T${\bf . \ }In our two state model the
cavity driving field drives transitions between lower and upper states, $%
\left| L\right\rangle $ and $\left| U\right\rangle $, in the $n$ and $n+1$
quanta manifolds respectively. \ The effective Rabi frequency of the
transition is given by 
\[
\Omega _{e}=\left| \beta \left\langle L\right| \hat{a}\left| U\right\rangle
\right| . 
\]
Under these conditions, the steady state population of $\left|
U\right\rangle $ is given by 
\[
\rho _{\text{exc}}=\frac{\Omega _{e}^{2}}{2\Omega _{e}^{2}+\Delta
_{e}^{2}+\Gamma _{U}^{2}} 
\]
where $\Delta _{e}$ is the detuning of the external cavity driving field
from the $\left| L\right\rangle -\left| U\right\rangle $ transition and $%
\Gamma _{U}$ the decay rate of state $\left| U\right\rangle $, assumed to
take population from $\left| U\right\rangle $ to $\left| L\right\rangle $. \
Note that we have ignored the decay rate from $\left| L\right\rangle $. \ We
denote the maximum value of $\rho _{\text{exc}}$ over all transitions from a
given lower state to all possible upper states in the $n$ quantum manifold
as $\rho _{\text{exc}}^{\left( n\right) }$. For ideal photon blockade, we
require $\rho _{\text{exc}}^{\left( 1\right) }\approx 0.5$ for the
transition from the ground state, $\left| L\right\rangle =\left| a,a,\ldots
,0\right\rangle =\left| G_{0}\right\rangle $ to the maximally coupled one
quantum dressed state $\left| G_{1}\right\rangle $. \ For the case that $%
\left| G_{1}\right\rangle =\left| D_{0}\right\rangle $, we note that $\Gamma
_{U}$ will be small, because $\left| D_{0}\right\rangle $ contains no
proportion of atomic state $\left| c\right\rangle $. \ Thus it is possible
to inject a single quantum of energy into the atom-cavity system for modest
values of $\beta $. \ Ideal blockade also requires that $\rho _{\text{exc}%
}^{\left( 2\right) }$ be negligible for transitions between $\left|
L\right\rangle =\left| G_{1}\right\rangle $ and states $\left|
U\right\rangle $ of the two quantum manifold.

We next consider the two atom, one quantum manifold of states. \ In this
case, the basis states are $\left| a,a,1\right\rangle $, $\left|
a,b,0\right\rangle $, $\left| a,c,0\right\rangle $, $\left|
b,a,0\right\rangle $ and $\left| c,a,0\right\rangle $ and the corresponding
block of $\widehat{{\cal H}}/\hbar $ can be written in matrix form as 
\begin{equation}
\frac{{\cal H}_{1}^{\left( 2\right) }}{\hbar }=\left[ 
\begin{array}{ccccc}
-i\Gamma _{\text{cav}} & 0 & g_{ac} & 0 & g_{ac} \\ 
0 & 0 & \Omega & 0 & 0 \\ 
g_{ac} & \Omega & -i\tilde{\Gamma}_{c} & 0 & 0 \\ 
0 & 0 & 0 & 0 & \Omega \\ 
g_{ac} & 0 & 0 & \Omega & -i\tilde{\Gamma}_{c}
\end{array}
\right]  \label{eq:MatrixForm:2Atom1Photon}
\end{equation}
Diagonalising ${\cal H}_{1}^{\left( 2\right) }/\hbar $ yields the
eigenvalues 
\begin{eqnarray*}
{\cal E}_{+2} &=&\left( -i\tilde{\Gamma}_{c}+\sqrt{-\tilde{\Gamma}%
_{c}^{2}+4\left( \Omega ^{2}+2g_{ac}^{2}\right) }\right) /2 \\
{\cal E}_{+1} &=&\left( -i\tilde{\Gamma}_{c}+\sqrt{-\tilde{\Gamma}%
_{c}^{2}+4\Omega ^{2}}\right) /2 \\
{\cal E}_{0} &=&0 \\
{\cal E}_{-1} &=&\left( -i\tilde{\Gamma}_{c}-\sqrt{-\tilde{\Gamma}%
_{c}^{2}+4\Omega ^{2}}\right) /2 \\
{\cal E}_{-2} &=&\left( -i\tilde{\Gamma}_{c}-\sqrt{-\tilde{\Gamma}%
_{c}^{2}+4\left( \Omega ^{2}+2g_{ac}^{2}\right) }\right) /2
\end{eqnarray*}
with associated dressed states $\left| D_{+2}\right\rangle $, $\left|
D_{+1}\right\rangle $, $\left| D_{0}\right\rangle $, $\left|
D_{-1}\right\rangle $ and $\left| D_{-2}\right\rangle $. The dressed state
energies are plotted in figure 2(b) for the same conditions as in figure
2(a). \ There are some important similarities between the spectrum of
eigenstates for the one atom and two atom cases. \ In each case there is a
state with zero energy, indicating that transitions from the zero to the one
quantum manifold are possible for a cavity driving field tuned to the cavity
resonance $\omega _{\text{cav}}$. \ The states which are anti-crossing in
each manifold are asymptotic to the lines$\ {\cal E}/\Omega =0$ and ${\cal E}%
/\Omega =\delta /\Omega $ with the point of closest approach being at $%
\delta /\Omega =0$. \ It is also important to realise that although there
are five distinct eigenstates only three of these eigenvalues will couple to
the ground state of the atom cavity system, i.e. the matrix element $%
\left\langle a,0\right| \hat{a}\left| D_{N}\right\rangle $ is non-zero only
for $N=0$, $\pm 2$. \ For this reason only the optically active states $%
\left| D_{+2}\right\rangle $, $\left| D_{0}\right\rangle $ and $\left|
D_{-2}\right\rangle $ are plotted in figure 2(b).

We have also solved the analogous three and four atom Hamiltonians and we
summarise our results for the eigenstates in each case as 
\begin{eqnarray}
{\cal E}_{+2} &=&\left( -i\tilde{\Gamma}_{c}+\sqrt{-\tilde{\Gamma}%
_{c}^{2}+4\left( \Omega ^{2}+Ng_{ac}^{2}\right) }\right) /2  \nonumber \\
{\cal E}_{+1} &=&\left( -i\tilde{\Gamma}_{c}+\sqrt{-\tilde{\Gamma}%
_{c}^{2}+4\Omega ^{2}}\right) /2  \nonumber \\
{\cal E}_{0} &=&0  \nonumber \\
{\cal E}_{-1} &=&\left( -i\tilde{\Gamma}_{c}-\sqrt{-\tilde{\Gamma}%
_{c}^{2}+4\Omega ^{2}}\right) /2  \nonumber \\
{\cal E}_{-2} &=&\left( -i\tilde{\Gamma}_{c}-\sqrt{-\tilde{\Gamma}%
_{c}^{2}+4\left( \Omega ^{2}+Ng_{ac}^{2}\right) }\right) /2
\label{eq:Eigenvalue:NAtom:1Quantum}
\end{eqnarray}
where $N=1,2,3,4$ indicates the number of atoms in the cavity. \ The
degeneracies of the eigenvalues presented in equations \ref
{eq:Eigenvalue:NAtom:1Quantum} are interesting to observe, namely the ${\cal %
E}_{+2}$, ${\cal E}_{0}$ and ${\cal E}_{-2}$ values are all non-degenerate,
whilst the ${\cal E}_{+1}$ and ${\cal E}_{-1}$ values are $\left( N-1\right) 
$ fold degenerate. (The latter values, ${\cal E}_{+1}$ and ${\cal E}_{-1}$
do not occur for $N=1$). \ The presence of the zero eigenvalue, ${\cal E}%
_{0} $, indicates that it is possible to inject one photon into the
atom-cavity system with a cavity driving field tuned to $\omega _{\text{cav}}
$.

Now we consider the two quanta manifold for a cavity containing one atom. \
This manifold is spanned by the states $\left| a,2\right\rangle $, $\left|
b,1\right\rangle $, $\left| c,1\right\rangle $ and $\left| d,0\right\rangle $
and the corresponding block of $\widehat{{\cal H}}/\hbar $\ in matrix form
is 
\[
\frac{{\cal H}_{2}^{\left( 1\right) }}{\hbar }=\left[ 
\begin{array}{cccc}
-2i\Gamma _{\text{cav}} & 0 & \sqrt{2}g_{ac} & 0 \\ 
0 & -i\Gamma _{\text{cav}} & \Omega & g_{bd} \\ 
\sqrt{2}g_{ac} & \Omega & -i\left( \Gamma _{\text{cav}}+\tilde{\Gamma}%
_{c}\right) & 0 \\ 
0 & g_{bd} & 0 & -i\widetilde{\Gamma }_{d}
\end{array}
\right] 
\]
In order to observe photon blockade in such a system, it is essential that
none of the eigenstates of ${\cal H}_{2}^{\left( 1\right) }/\hbar $ are
resonantly coupled by the cavity driving field to the occupied states of the
one quantum manifold. \ To achieve this, we require $\rho _{\text{exc}%
}^{\left( 2\right) }\ll 0.5$, the smaller $\rho _{\text{exc}}^{\left(
2\right) }$ the greater the degree of photon blockade. \ A plot showing the
eigenenergies of this system with associated linewidths, $\Gamma _{x}$ where 
$x$ is the dressed state under consideration, is presented in figures 3(a)
and 3(b). \ In each case $\Delta /\Omega =2$, $\Gamma _{c}/\Omega =\Gamma
_{d}/\Omega =0.1$, $\Gamma _{\text{cav}}/\Omega =0.01,$ $g_{ac}/\Omega =1$
and in figure 3(a) $g_{bd}/\Omega =1$ and the energies are plotted as a
function of $\delta /\Omega ,$ whereas in figure 3(b) $\delta /\Omega =0$
and the energies are plotted as a function of $g_{bd}/\Omega $. \ The
important features to recognise from these two traces is the shift of the
smallest magnitude energy state from zero, indicating (as highlighted in 
\cite{bib:Werner1999} and \cite{bib:Rebic1999}) that photon blockade will
indeed be possible in the one atom case.

Similarly, for the two atom case, the basis states of the two quanta
manifold are $\left| a,a,2\right\rangle ,$ $\left| a,b,1\right\rangle $, $%
\left| a,c,1\right\rangle $, $\left| b,a,1\right\rangle $, $\left|
c,a,1\right\rangle $, $\left| a,d,0\right\rangle $, $\left|
b,b,0\right\rangle $, $\left| b,c,0\right\rangle $, $\left|
c,b,0\right\rangle $, $\left| c,c,0\right\rangle $,and $\left|
d,a,0\right\rangle $ where the notation is $\left| \text{atom 1, atom 2,
cavity field}\right\rangle $. In matrix form, the corresponding block of the
Hamiltonian is 
\[
\frac{{\cal H}_{2}^{\left( 2\right) }}{\hbar }=\left[ 
\begin{array}{ccccccccccc}
-2i\Gamma _{\text{cav}} & 0 & \sqrt{2}g_{ac} & 0 & \sqrt{2}g_{ac} & 0 & 0 & 0
& 0 & 0 & 0 \\ 
0 & -i\Gamma _{\text{cav}} & \Omega & 0 & 0 & g_{bd} & 0 & 0 & g_{ac} & 0 & 0
\\ 
\sqrt{2}g_{ac} & \Omega & -i\Gamma _{t} & 0 & 0 & 0 & 0 & 0 & 0 & g_{ac} & 0
\\ 
0 & 0 & 0 & -i\Gamma _{\text{cav}} & \Omega & 0 & 0 & g_{ac} & 0 & 0 & g_{bd}
\\ 
\sqrt{2}g_{ac} & 0 & 0 & \Omega & -i\Gamma _{t} & 0 & 0 & 0 & 0 & g_{ac} & 0
\\ 
0 & g_{bd} & 0 & 0 & 0 & -i\widetilde{\Gamma }_{d} & 0 & 0 & 0 & 0 & 0 \\ 
0 & 0 & 0 & 0 & 0 & 0 & 0 & \Omega & \Omega & 0 & 0 \\ 
0 & 0 & 0 & g_{ac} & 0 & 0 & \Omega & -i\widetilde{\Gamma }_{c} & 0 & \Omega
& 0 \\ 
0 & g_{ac} & 0 & 0 & 0 & 0 & \Omega & 0 & -i\widetilde{\Gamma }_{c} & \Omega
& 0 \\ 
0 & 0 & g_{ac} & 0 & g_{ac} & 0 & 0 & \Omega & \Omega & -i\Gamma _{t} & 0 \\ 
0 & 0 & 0 & g_{bd} & 0 & 0 & 0 & 0 & 0 & 0 & -i\widetilde{\Gamma }_{d}
\end{array}
\right] 
\]
where $\Gamma _{t}=\Gamma _{\text{cav}}+\widetilde{\Gamma }_{c}$.

The energy eigenvalues of ${\cal H}_{2}^{\left( 2\right) }/\hbar $ are
presented in figures 4(a) and 4(b). \ As in the two atom, single quantum
case there are optically inactive states, and these have been removed from
the figures. \ Figure 4(a) was generated for $\Delta /\Omega =2$, $\delta
/\Omega =0$, $\Gamma _{c}/\Omega =\Gamma _{d}/\Omega =0.1$, $\Gamma _{\text{%
cav}}/\Omega =0.01$ and $g_{ac}/\Omega =1$. \ As can be seen, for this case
there is a zero eigenvalue so we would expect that there would be resonant
coupling from the single quantum manifold to the two quantum manifold by a
cavity driving field tuned to $\omega _{\text{cav}}$ and hence, photon
blockade would not be observed in this case. \ However in figure 4(b) we
illustrate the effect of introducing a small mutual detuning, $\delta ,$
from state $\left| c\right\rangle $. \ The parameters used were $\Delta
/\Omega =2$, $\delta /\Omega =0.5$, $\Gamma _{c}/\Omega =\Gamma _{d}/\Omega
=0.1$, $\Gamma _{\text{cav}}/\Omega =0.01$ and $g_{ac}/\Omega =1$. \ It is
important to observe the shift of the smallest magnitude eigenvalue from
zero. \ Although this shift is less than for the corresponding single atom
case, it is still feasible to consider building a cavity to observe this
photon blockade. \ This point will be elaborated on in the conclusions\cite
{NOTE:Rebic}.

\section{Photon blockade in a realistic system}

Of critical importance to the experimental observation of photon blockade in
atomic systems, is the question of how to trap single or small numbers of
atoms within a small, high finesse optical cavity. \ We know of no
demonstration of continuous trapping, however recent experiments conducted
by Hood {\it et al}\cite{bib:Hood1998} and M\"{u}nstermann {\it et al}\cite
{bib:Munstermann} have shown that it is possible to have a cavity with
extremely low atom fluxes passing through it. \ This was achieved in \cite
{bib:Hood1998} by allowing atoms to fall from a leaky MOT into the cavity
and in \cite{bib:Munstermann} by directing atoms out of a MOT and into the
cavity. \ One would expect such atomic ejections to be stochastic in nature
and as a consequence, the probability of having a certain number of atoms in
the cavity should follow Poissonian statistics. \ In order to build a device
which uses photon blockade, it will therefore be necessary to ensure that
significant photon blockade will be observed over as wide a range of atomic
numbers as possible, otherwise the photon blockade could be lost and the
performance of the device degraded.

A complication in the experimental realisation of photon blockade is that
the parameters chosen in the theoretical plots shown above were assumed to
be independent variables. \ In general this will not be the case unless
great care is taken in the preparation of an experiment. \ Specifically, for
transitions in an alkali vapour which might realise the N configuration, one
would expect the dipole moments of all relevant transitions to be of the
same order, with the atom-cavity coupling set solely by the cavity volume, a
parameter shared by both the $\left| a\right\rangle -\left| c\right\rangle $
and $\left| b\right\rangle -\left| d\right\rangle $ transitions. \ One would
therefore expect $g_{ac}\sim g_{bd}=g$. \ Also, because of the shared
cavity, the detuning parameters are not \ independent, so if we assume that $%
\omega _{ca}-\omega _{db}=\delta _{\omega }$ then we find that $\Delta
=\delta +\delta _{\omega }$. \ These extra considerations will be included
in the analysis to follow, which concentrates on experimentally realisable
effects rather than the general theoretical demonstration presented above. \
We start by investigating photon blockade using the parameters $\Gamma _{%
\text{cav}}/\Omega =0.01$, $\Gamma _{c}/\Omega =\Gamma _{d}/\Omega =0.1$, $%
\delta _{w}/\Omega =0$ and $\beta /\Omega =1$. \ For these values $\rho _{%
\text{exc}}^{\left( 1\right) }\approx 0.5$, as is expected due to the
presence of the strongly absorbing state at the cavity resonance.

In figure 5 (a) we show a pseudo-colour plot of $\rho _{\text{exc}}^{\left(
2\right) }$ as a function of $\delta /\Omega $ and $g/\Omega $ for the one
atom two quanta case, where colour is indicative of the value of $\rho _{%
\text{exc}}^{\left( 2\right) }$, blue being 0 and red being 0.5. \ The graph
clearly shows $\rho _{\text{exc}}^{\left( 2\right) }$ decreasing
monotonically with $g/\Omega $, indicating the effectiveness of photon
blockade correspondingly increasing. \ This is expected, as a large $g$ will
give rise to a highly nonlinear system. \ Also note that $\rho _{\text{exc}%
}^{\left( 2\right) }$ increases as $\left| \delta /\Omega \right| $
increases, indicating that the photon blockade is a resonance phenomenon.

In figure 5(b) we show the analogous $\rho _{\text{exc}}^{\left( 2\right) }$
plot for two atoms. \ In accord with the preliminary results which suggested
that photon blockade was not possible in the multi-atom system\cite
{bib:Imamoglu1998,bib:Grangier1998,bib:Rebic1999} we find $\rho _{\text{exc}%
}^{\left( 2\right) }\lesssim 0.5$ in the vicinity of $\delta /\Omega =0$,
implying that photon blockade will not be observable for these choices of
parameters. \ However, by increasing the mutual detuning and employing a
modest atom-cavity coupling, regions of strong photon blockade are
observable, typified by the minimum recorded value on figure 5(b) of $\rho _{%
\text{exc}}^{\left( 2\right) }=0.0087$.

The three atom case is shown in figure 5(c) for the same parameter regime as
the one and two atom cases. \ The form of $\rho _{\text{exc}}^{\left(
2\right) }$ is similar to that of the two atom case, with the same overall
structure, but narrower regions where photon blockade should be observable,
this is shown by the smaller blue region of figure 5(c) than figure 5(b). \
Again very low values of $\rho _{\text{exc}}^{\left( 2\right) }$ were
obtained, with a minimum recorded value of $\rho _{\text{exc}}^{\left(
2\right) }=0.0070$. \ 

To use photon blockade as a tool for new quantum devices it is necessary to
identify real systems in which these effects may be observed. \ As an
example of what may be achieved with the current state of the art, we take
parameters from a recent experimental paper \cite{bib:Hood1998} and make
some minor assumptions about how they might apply to atoms falling through a
high finesse cavity. \ It should certainly be possible to achieve a coupling
field Rabi frequency of $\Omega =10$ MHz, external field-cavity coupling
strength of $\beta =1$MHz and Hood {\it et al} achieve $g=120$ MHz and $%
\Gamma _{\text{cav}}=40$ MHz. \ For a system based on transitions in the $%
^{87}$Rb D$_{2}$ line \cite{bib:ChenThesis} we may assume $\delta _{\omega
}=6600$MHz and $\Gamma _{c}=\Gamma _{d}=17.8$ MHz. \ We ignore the Zeeman
magnetic sublevels since these form a simple, effective three level $\Lambda 
$ system (for states $\left| a\right\rangle $, $\left| b\right\rangle $ and $%
\left| c\right\rangle $) for the situation under consideration \cite
{bib:Li1995}. \ Assuming equal rates of radiative decay, these translate
into our system as $g/\Omega =12$, $\Gamma _{c}/\Omega =\Gamma _{d}/\Omega
=1.78$, $\Gamma _{\text{cav}}/\Omega =4$, $\delta _{\omega }/\Omega =660$
and $\beta /\Omega =0.3$. \ In figure 6(a) we present a plot of $\rho _{%
\text{exc}}^{\left( 1\right) }$ as a function of $g/\Omega $ and $\delta
/\Omega $ for one atom in the cavity, with the plots showing $\rho _{\text{%
exc}}^{\left( 2\right) }$ in (b) and (c) for one and two atoms respectively.

There are some important features to note in figures 6. \ In 6(a) $\rho _{%
\text{exc}}^{\left( 1\right) }$ increases monotonically with $g/\Omega $
and, for the scale used, is independant of $\delta /\Omega $. \ The value of 
$\rho _{\text{exc}}^{\left( 1\right) }$ is qualitatively very similar for
one, two and three atoms, with the values of $\rho _{\text{exc}}^{\left(
1\right) }$ slightly increasing as the number of atoms increases. \ As an
example, for $g/\Omega =12$, $\rho _{\text{exc}}^{\left( 1\right) }=0.3099$, 
$0.3824$ and $0.4148$ for one, two and three atoms respectively. \ Traces
6(b) and (c) show $\rho _{\text{exc}}^{\left( 2\right) }$ for the one and
two atom cases respectively. \ The value of $\rho _{\text{exc}}^{\left(
2\right) }$ for the one atom case is very small across the entire parameter
space, indicating that photon blockade should be easy to observe for a
single atom. Of interest is the resonance in the vicinity of $\delta =\delta
_{w}$ which will be discussed below. \ There is an extra resonance in the
vicinity of $\delta =0$ for the one atom case, although this is not present
to the same extent in the two and three atom cases. \ For two atoms, and
also for three atoms (not shown), the overall values (away from the
resonances) of $\rho _{\text{exc}}^{\left( 2\right) }$ appear to increase
with the number of atoms and are much larger than for the one atom case.
Qualitatively, $\rho _{\text{exc}}^{\left( 2\right) }$ for two and three
atoms are extremely similar, with maximum values along the line $g/\Omega
=12 $ of $\rho _{\text{exc}}^{\left( 2\right) }=0.2244$ and $0.3092$ for two
and three atoms respectively. \ These observations would appear to agree
with the intuitive idea that photon blockade would be more difficult in
multi-atom systems and that there is a qualitative difference between the
single atom case and multi-atom cases.

The results for $\rho _{\text{exc}}^{\left( 2\right) }$ in the vicinity of $%
\delta /\Omega =-660$ are shown in figure 7. \ The significance of this
region is that we have $\delta =-\delta _{w}$ so that the cavity is now
resonant with the $\left| b\right\rangle -\left| d\right\rangle $
transition. \ To our knowledge, this situation has not been previously
explored and the consequences it has for photon blockade are significant. \
In figure 7 we present $\rho _{\text{exc}}^{\left( 2\right) }$ for one, two
and three atoms in plots (a), (b) and (c) respectively. \ The behaviour of $%
\rho _{\text{exc}}^{\left( 1\right) }$ shows no resonance phenomena and is
described above, it is only by considering $\rho _{\text{exc}}^{\left(
2\right) }$ that the resonance is observed. \ In 7(a) we see generally small
values for $\rho _{\text{exc}}^{\left( 2\right) }$ indicating that photon
blockade should be observable. \ With the exception of the `shelf' for $%
g/\Omega \lesssim 1$, $\rho _{\text{exc}}^{\left( 2\right) }$ has the a
similar chevron shape to that observed in the ideal case, shown in figure
5(a), although with very much smaller values. \ In figures 7(b) and (c), we
observe a roughly triangular region of low $\rho _{\text{exc}}^{\left(
2\right) }$, superimposed on the background of $\rho _{\text{exc}}^{\left(
2\right) }$ noted earlier. \ Values of $\rho _{\text{exc}}^{\left( 2\right)
}<0.01$ are present where $\rho _{\text{exc}}^{\left( 1\right) }>0.35$ for
both the two and three atom cases. \ This suggests that the non-linearity of
the system is very large about this resonance. \ Further investigations of
this resonance will be done with increasing number of atoms to show how
robust the photon blockade will be as any relaxation of the requirement for
low number of atoms will enhance the prospects for experimental verification
of this effect.

\section{Photon blockade with an off resonant cavity}

The discussions above have used a system where the cavity is driven
resonantly by the external field. \ This has a significant drawback in
trying to realise a continuously operating device inasmuch as when there are
no atoms in the cavity, the cavity will absorb photons from the external
field. \ To overcome this using the on-resonance configuration, one must run
the experiment in a pulsed mode to ensure that there are no photons in the
cavity prior to atoms entering the cavity. \ There is, however, another
alternative. \ This occurs when the cavity driving field is not tuned to the
cavity resonance, but instead is tuned to a side resonance of the one atom
dressed system. In this case there will be strong coupling between the
external driving field and the cavity when there is one atom in the cavity
and no coupling when there are zero atoms in the cavity. \ By studying the
form of the dressed state eigenvalues given in equation \ref
{eq:Eigenvalue:NAtom:1Quantum} it is clear that there is a dependence of the
eigenvalue with the number of atoms so that it should be possible to choose $%
g_{ac}$ such that the one-quantum manifold is only coupled into when there
is only one atom in the cavity. \ These conditions are best met for the case
that both $\delta $ and $\Omega $ are small compared to $g_{ac}$ to ensure
the maximum shift in eigenvalues with number of atoms.

\section{Conclusions}

We have presented detailed calculations which show the parameter space where
photon blockade in one, two and three atom systems should be observable. \
By considering transitions on the $^{87}$Rb D$_{2}$ line, we have suggested
that photon blockade should be observable using current state of the art
technology, such as has been recently demonstrated \cite{bib:Hood1998}.

We have pointed out that strong photon blockade should be possible when the
cavity resonance is tuned to the $\left| b\right\rangle -\left|
d\right\rangle $ transition, despite a large mutual detuning, $\delta $. \
This resonance may relax the requirements on the energy levels of the atomic
species under consideration.

We gratefully acknowledge financial support from the EPSRC and useful
discussions with Ole Steuernagel (University of Hertfordshire), Derek
Richards (The Open University), Danny Segal and Almut Beige (Imperial
College).

\section{Figures}

Figure 1: Energy level diagram for the four level N system. \ The atoms are
labelled in order of increasing energy as $\left| a\right\rangle $, $\left|
b\right\rangle $, $\left| c\right\rangle $, and $\left| d\right\rangle $
with energies $\hbar \omega _{a}$, $\hbar \omega _{b}$, $\hbar \omega _{c}$
and $\hbar \omega _{d}$ respectively. \ A strong classical coupling field
with frequency $\omega _{\text{class}}$ and Rabi frequency $\Omega $ is
applied to the $\left| b\right\rangle -\left| c\right\rangle $ transition
and detuned from it by an amount $\delta =\omega _{cb}-\omega _{\text{class}%
} $. \ The atoms are placed in a cavity with resonance frequency $\omega _{%
\text{cav}}$ which is detuned from the $\left| a\right\rangle -\left|
c\right\rangle $ transition by $\delta $ and from the $\left| b\right\rangle
-\left| d\right\rangle $ transition by $\Delta =\omega _{db}-\omega _{\text{%
cav}}$. \ The cavity is driven resonantly by an external classical driving
field with external field-cavity coupling strength $\beta $ and frequency $%
\omega _{\text{cav}}$.

Figure 2: Eigenvalues of the one quantum manifold as a function of the
scaled mutual detuning $\delta /\Omega $ for one and two atoms in the
cavity. \ In each case the parameters used were $g_{ac}/\Omega =1$, $\Gamma
_{\text{cav}}/\Omega =0.01$ and $\Gamma _{c}/\Omega =0.1$. \ In figure 2(a)
the spectrum for a single atom interacting with a single cavity photon
traces out the well known Mollow triplet\cite{bib:Mollow1969}. \ In figure
2(b) we present the analogous trace for two atoms instead of one where the
optically inactive dressed states, $\left| D_{+1}\right\rangle $ and $\left|
D_{-1}\right\rangle $ have been removed.

Figure 3: Eigenvalues of the two quanta manifold for a single atom. \
Parameters used were the same as those in figure 2, but with the addition of 
$\Gamma _{d}/\Omega =0.1$ and $\Delta /\Omega =2$. \ Figure 3(a) has $%
g_{bd}/\Omega =1$ and the energy eigenvalues are plotted as a function of
scaled mutual detuning, whilst figure 3(b) has $\delta /\Omega =0$ and the
energy eigenvalues are plotted as a function of $g_{bd}/\Omega $.

Figure 4: Eigenvalues of the two quanta manifold for two atoms in the cavity
with the parameters $\Delta /\Omega =2$, $\Gamma _{c}/\Omega =\Gamma
_{d}/\Omega =0.1$, $\Gamma _{\text{cav}}/\Omega =0.01$ and $g_{ac}/\Omega =1$%
, as a function of $g_{bd}/\Omega $. In 4(a) the mutual detuning $\delta
/\Omega =0$ and no photon blockade is observed, in 4(b) a small mutual
detuning of $\delta /\Omega =0.5$ is applied and the photon blockade is
restored for a cavity driving field tuned to $\omega _{\text{cav}}$. \ Note
that the optically inactive states have been removed in these traces also.

Figure 5: $\rho _{\text{exc}}^{\left( 2\right) }$ plotted as a function of
scaled detuning $\left( \delta /\Omega \right) $ and scaled coupling $\left(
g/\Omega \right) $\ for one atom (a), two atoms (b) and three atoms (c). \
The colour of the plot shows the value of $\rho _{\text{exc}}^{\left(
2\right) }$ with blue being zero, red 0.5. \ The parameters chosen for these
figures were $\Gamma _{\text{cav}}/\Omega =0.01$, $\Gamma _{c}/\Omega
=\Gamma _{d}/\Omega =0.1$, $\delta _{w}/\Omega =0$ and $\beta /\Omega =1$. \
Note that over this paramter range, $\rho _{\text{exc}}^{\left( 1\right)
}=0.5$.

Figure 6: \ Pseudo-colour plots of $\rho _{\text{exc}}^{\left( 1\right) }$
and $\rho _{\text{exc}}^{\left( 2\right) }$. The parameters chosen were
those which could be expected in a realistic experiment involving $^{87}$Rb
atoms. \ These parameters were $\Gamma _{c}/\Omega =\Gamma _{d}/\Omega =1.78$%
, $\Gamma _{\text{cav}}/\Omega =4$, $\delta _{\omega }/\Omega =660$ and $%
\beta /\Omega =0.3$. \ Figure (a) corresponds to $\rho _{\text{exc}}^{\left(
1\right) }$ for one atom, whilst (b) and (c) correspond to $\rho _{\text{exc}%
}^{\left( 2\right) }$ for one and two atoms respectively. \ Note that the
colour scales for each figure vary according to the maximum values of $\rho
_{\text{exc}}$.

Figure 7: Pseudo-colour plots of $\rho _{\text{exc}}^{\left( 2\right) }$ in
the vicinity of the resonance $\delta =-\delta _{w}$ for one (a), two (b)
and three (c) atoms. \ The parameters used were the same as in figure 6.

\end{document}